\documentstyle[prl,aps]{revtex}
\begin{document}
\author{Jian-Qi Shen\footnote{E-mail address: jqshen@coer.zju.edu.cn}}
\address{Zhejiang Institute of Modern Physics and Department of
Physics, Zhejiang University, Hangzhou 310027, P.R. China}
\date{\today}
\title{The Pegg-Barnett oscillator and its supersymmetric generalization}
\maketitle

\begin{abstract}
The oscillator algebra of Pegg-Barnett (P-B) oscillator with a
finite-dimensional number-state space is investigated in this
note. It is shown that the Pegg-Barnett oscillator possesses the
su($n$) Lie algebraic structure. Additionally, we suggest a
so-called supersymmetric P-B oscillator and discuss the related
topics such as the algebraic structure and particle occupation
number of supersymmetric P-B oscillator.
\end{abstract}

It is well known that the usual mathematical model of the monomode
quantized electromagnetic field is the harmonic oscillator with an
infinite-dimensional number-state space, the commuting relation of
which is $[a, a^{\dagger}]={\mathcal I}$ with $a$ and
$a^{\dagger}$ being respectively the annihilation and creation
operators of single-mode photon fields. Due to the permutation
invariance in the trace of the product of two matrices
(operators), {\it i.e.}, ${\rm tr} (aa^{\dagger})={\rm tr}
(a^{\dagger}a)$, it follows directly that the trace of commutator
is vanishing, {\it i.e.}, ${\rm tr}[a, a^{\dagger}]=0$, which,
however, contradicts the fact that the unit matrix ${\mathcal I}$
possesses a nonvanishing trace, namely, ${\rm tr}{\mathcal I}\neq
0$. The above brief discussion shows that there exists no
finite-dimensional representations of Heisenberg algebra
(non-semisimple Lie algebra). This, therefore, means that we
should consider the oscillator algebra with finite-dimensional
state spaces, which is a semisimple algebraic extension of Bosonic
oscillator algebra. On the other hand, in an attempt to
investigate the number-phase uncertainty relations of the maser
and squeezed state in quantum optics, physicists meet, however,
with difficulties arising from a fact that the classical
observable phase of light {\it unexpectedly} has no corresponding
Hermitian operator counterpart (quantum optical
phase)\cite{Louisell,Susskind,Carruthers}. So, several problems we
encountered are as follows: (i) the exponential-form operator
$\exp [i\hat{\phi}]$ (with $\hat{\phi}$ being the phase operator)
is not unitary; (ii) the number-state expectation value of Dirac's
quantum relation $[\hat{\phi}, \hat{N}]=-i$ (with $\hat{N}$ being
the occupation-number operator of photon fields) is even zero,
{\it i.e.}, $\langle n |[\hat{\phi}, \hat{N}]| n\rangle=0$; (iii)
the number-phase uncertainty relation $\Delta N\Delta \phi\geq
\frac{1}{2}$ would imply that a well-defined number state would
actually have a phase uncertainty of greater than $2
\pi$\cite{Pegg}. In order to overcome these difficulties, Pegg and
Barnett suggested an alternative, and physically
indistinguishable, mathematical model of the single-mode field
involving a finite but arbitrarily large state space\cite{Pegg},
in which they defined a phase state as follows
 \begin{equation}
|\theta\rangle=\lim_{s\rightarrow\infty}(s+1)^{-\frac{1}{2}}\sum^{s}_{n=0}\exp
(in\theta)|n\rangle,
\end{equation}
where $|n\rangle$ are the $s+1$ number states, which span an
$(s+1)$ -dimensional state space. This, therefore, means that the
state space $\{|n\rangle\}$ with $0 \leq n\leq s$ has a finitely
upper level ($|s\rangle$) and the maximum occupation number of
particles is $s$ rather than infinity. In their new quantization
formulation, the dimension of number state space is allowed to
tend to infinity after physically measurable results are
calculated\cite{Pegg}. Pegg and Barnett showed that this approach
and the conventional infinite state space are physically
indistinguishable. However, this method has the additional
advantage of being able to incorporate a well-behaved Hermitian
phase operator within the formalism. The resulting number-phase
commutator in Pegg-Barnett approach does not lead to any
inconsistencies yet satisfies the condition for
Poisson-bracket-commutator correspondence. It was shown that
Pegg-Barnett approach has several advantages over the conventional
Susskind-Glogower formulation\cite{Susskind}. For example, the
Pegg-Barnett phase operator is consistent with the vacuum being a
state of random phase, while the Susskind-Glogower phase operator
does not demonstrate such property of the vacuum\cite{Pegg}.
Pegg-Barnett formulation is useful for treating the problems of
atomic coherent population trapping (CPT) and electromagnetically
induced transparency (EIT)\cite{Purdy}.

In this note we will further consider the Pegg-Barnett harmonic
oscillator that involves a finitely large state space, and show
that it possesses the su($n$) Lie algebraic structures. Based on
this consideration, we will generalize the Pegg-Barnett oscillator
to a supersymmetric one.

The quantum harmonic oscillator possessing an infinite-dimensional
number-state space ({\it i.e.}, the maximum occupation number $s$
tends to infinity) can well model the Bosonic fields. Taking
account of the Pegg-Barnett (P-B) harmonic oscillator means that
the non-semisimple Lie algebra should be generalized to the
semisimple one, namely, we will replace the familiar quantum
commutator $[a, a^{\dagger}]={\mathcal I}$ with the new $[a,
a^{\dagger}]={\mathcal A}$ (${\mathcal A}$ will be defined in what
follows). For a preliminary consideration, we take into account
the case $s=1$, where the matrix representation of the
annihilation (creation) operators and ${\mathcal A}$ of the fields
are of the form (in the number-state basis set)
\begin{equation}
a=\left({\begin{array}{cccc}
{0 } &{ 1 } \\
{0} &  { 0}  \\
 \end{array} }
 \right),                 \quad         a^{\dagger}=\left({\begin{array}{cccc}
{0} & {0} \\
{1} & {0}  \\
 \end{array}}
 \right),                     \quad         {\mathcal A}=\left({\begin{array}{cccc}
{1} & {0} \\
{0} & {-1}  \\
 \end{array}}
 \right).
 \label{eq1}
\end{equation}
It is apparently seen that the operators $a$, $a^{\dagger}$ and
${\mathcal A}$ satisfy an sl($2$) algebraic commuting relations.
Here one can readily verify that
$a=\frac{\sigma_{1}+i\sigma_{2}}{2}$,
$a^{\dagger}=\frac{\sigma_{1}-i\sigma_{2}}{2}$ and ${\mathcal
A}=\sigma_{3}$, where $\sigma_{i}$'s ($i=1,2,3$) are Pauli's
matrices. It follows from (\ref{eq1}) that
$aa^{\dagger}+a^{\dagger}a={\mathcal I}$. It is apparently seen
that the algebraic generators of su(2) Lie algebra can be
constructed in terms of the matrices in (\ref{eq1}). This,
therefore, implies that the P-B harmonic oscillator with $s=1$
corresponds to the Fermionic fields and possesses the su($2$) Lie
algebraic structure.

In what follows we will study the algebraic structures of P-B
harmonic oscillators with arbitrary occupation number $s$.  As
another illustrative example, here we will take into consideration
the case of $s=2$, the matrix representation of $a$, $a^{\dagger}$
and ${\mathcal A}$ of which are written (in the number-state basis
set)
\begin{equation}
a=\left({\begin{array}{cccc}
{0} & {1} & {0} \\
{0} &   {0} & {\sqrt{2}}  \\
 {0} &  {0} &  {0}         \\
 \end{array}}
 \right),                 \quad          a^{\dagger}=\left({\begin{array}{cccc}
{0}  & {0} & {0} \\
{1} &   {0} & {0}  \\
 {0} &  {\sqrt{2}} &  {0}    \\
 \end{array} }
 \right),                 \quad          {\mathcal A}=\left({\begin{array}{cccc}
{1}  & {0} & {0} \\
{0} &   {1} & {0}  \\
 {0} &  {0} &  {-2}      \\
 \end{array}  }
 \right).
 \label{eq2}
\end{equation}
Calculation of the commutators among the Lie algebraic generators
of the P-B harmonic oscillator with $s=2$ yields
\begin{eqnarray}
\left[a,  {\mathcal
A}\right]&=&3\sqrt{2}\left({\begin{array}{cccc}
{0}  & {0} & {0} \\
{0}&  {0} &{-1}  \\
 {0} &  {0} &  {0}    \\
 \end{array}  }
 \right)=3\sqrt{2}{\mathcal M},      \quad          \quad     [a^{\dagger},  {\mathcal A}]
 =-3\sqrt{2}\left({\begin{array}{cccc}
{0}  & {0} & {0} \\
{0}&  {0} &{0}  \\
 {0} &  {-1} &  {0}      \\
 \end{array}  }
 \right)=-3\sqrt{2}{\mathcal M}^{\dagger},             \nonumber \\
\left[{\mathcal M}, {\mathcal M}^{\dagger}\right]
&=&-\left({\begin{array}{cccc}
{0}  & {0} & {0} \\
{0}&  {-1} &{0}  \\
{0} &  {0}&  {1}      \\
\end{array} }
\right)=-{\mathcal K},                    \quad          \quad
                         \left[a,  {\mathcal M}\right]=-\left({\begin{array}{cccc}
{0}  & {0} & {1} \\
{0}&  {0} &{0}  \\
 {0} &  {0} &  {0}    \\
 \end{array}}
 \right)=-{\mathcal F},
                \nonumber \\
 \left[a^{\dagger}, {\mathcal M}\right]&=&-\sqrt{2}{\mathcal K},
        \quad      \left[a, {\mathcal M}^{\dagger}\right]=\sqrt{2}{\mathcal K},
           \quad       \left[a^{\dagger}, {\mathcal M}^{\dagger}\right]={\mathcal F}^{\dagger},
                 \quad     [{\mathcal K}, {\mathcal F}]=-{\mathcal
                 F},  \quad      [{\mathcal K}, {\mathcal F}^{\dagger}]={\mathcal
                 F}^{\dagger}, \quad  ...    \label{eq3}
\end{eqnarray}
Further calculation shows that the algebraic generators $a,
a^{\dagger}, {\mathcal A}, {\mathcal M}, {\mathcal M}^{\dagger},
{\mathcal K}, {\mathcal F}, {\mathcal F}^{\dagger}$ form the
sl($3$) algebra. The eight Gell-Mann matrices can therefore be
constructed in terms of them, {\it i.e.},
\begin{eqnarray}
\lambda_{1}&=&a+a^{\dagger}+\sqrt{2}({\mathcal M}+{\mathcal
M}^{\dagger}),   \quad
\lambda_{2}=i[a^{\dagger}-a+\sqrt{2}({\mathcal
M}^{\dagger}-{\mathcal M})], \quad     \lambda_{3}={\mathcal
A}+2{\mathcal K},  \quad    \lambda_{4}={\mathcal F}+{\mathcal
F}^{\dagger},
                 \nonumber \\
    \lambda_{5}&=&i({\mathcal
F}^{\dagger}-{\mathcal F}),   \quad \lambda_{6}=-({\mathcal
M}+{\mathcal M}^{\dagger}), \quad \lambda_{7}=-i({\mathcal
M}^{\dagger}-{\mathcal M}),     \quad
\lambda_{8}=\frac{1}{\sqrt{3}}\lambda_{8}.
\end{eqnarray}
Thus we show that the P-B harmonic oscillator with $s=2$ possesses
the su($3$) Lie algebraic structure.

For the P-B harmonic oscillator with a finite but arbitrarily
large state space of $s+1$ dimensions, the matrix representation
(in the number-state basis set) of the operators $a$,
$a^{\dagger}$ and ${\mathcal A}$ takes the following form
\begin{equation}
a_{mn}=\sqrt{n}\delta_{m, n-1},  \quad
a^{\dagger}_{mn}=\sqrt{n+1}\delta_{m, n+1},   \quad    {\mathcal
A}_{mn}=\delta_{mn}-(s+1)\delta_{ms}\delta_{ns},
\end{equation}
where the subscript $m, n$ (which run from $0$ to $s$ only) denote
the matrix row-column indices. The remaining generators ${\mathcal
M}, {\mathcal M}^{\dagger}, {\mathcal K}, {\mathcal F}, {\mathcal
F}^{\dagger}, ...$ can be obtained as follows ($0\leq m, n\leq
s$):
\begin{eqnarray}
\left[a, {\mathcal A}\right]_{mn}&=&(s+1)\sqrt{s}(-\delta_{m+1,
s}\delta_{n
s})=(s+1)\sqrt{s}{\mathcal M}_{mn},   \nonumber \\
\left[a^{\dagger}, {\mathcal
A}\right]_{mn}&=&-(s+1)\sqrt{s}(-\delta_{m s}\delta_{n+1,
s})=-(s+1)\sqrt{s}{\mathcal M}^{\dagger}_{mn},
              \nonumber \\
\left[{\mathcal M}, {\mathcal
M}^{\dagger}\right]_{mn}&=&-(\delta_{ms}\delta_{ns}-\delta_{m+1,s}\delta_{n+1,s})=-{\mathcal
K}_{mn},
\nonumber \\
\left[ {\mathcal A}, {\mathcal M}\right]&=&(1+s){\mathcal M},
\quad           \left[ {\mathcal A}, {\mathcal
M}^{\dagger}\right]=-(1+s){\mathcal M}^{\dagger},
\nonumber \\
\left[a, {\mathcal M}\right]_{mn}&=&-\sqrt{s-1}\delta_{m+1,
s-1}\delta_{ns}=-\sqrt{s-1}{\mathcal F}_{mn},
\nonumber \\
\left[a^{\dagger}, {\mathcal
M}^{\dagger}\right]_{mn}&=&\sqrt{s-1}\delta_{ms}\delta_{n+1,
s-1}=\sqrt{s-1}{\mathcal F}^{\dagger}_{mn}, \nonumber  \\
\left[{\mathcal K}, {\mathcal F}\right]&=&-{\mathcal F}, \quad
\left[{\mathcal K}, {\mathcal F}^{\dagger}\right]={\mathcal
F}^{\dagger},   \quad    \left[{\mathcal M}, {\mathcal
K}\right]=2{\mathcal M},  \quad             \left[{\mathcal
M}^{\dagger}, {\mathcal K}\right]=-2{\mathcal M}^{\dagger}, \quad
...                      \label{eq4}
\end{eqnarray}
For the case of $s=2$, it has been shown above that Hermitian
operators (such as the eight Gell-Mann matrices) can be
constructed in terms of $a, a^{\dagger}, {\mathcal A}, {\mathcal
M}, {\mathcal M}^{\dagger}, {\mathcal K}, {\mathcal F}, {\mathcal
F}^{\dagger}$. Likewise, here the Hermitian operators (generators)
of Lie algebra can also be obtained via the linear combination of
the above generators (\ref{eq4}). If ${\mathcal G}$ represents the
linear combination of the Hermitian operators, and consequently
${\mathcal G}={\mathcal G}^{\dagger}$, then the exponential-form
group element operator $U=\exp (i{\mathcal G})$ is unitary.
Besides, since $a$, $a^{\dagger}$ and ${\mathcal A}$ are
traceless, all the generators derived by the commutators in
(\ref{eq4}) (and hence ${\mathcal G}$) are also traceless due to
the cyclic invariance in the trace of matrices product. Thus the
determinant of the group element $U$ is unit, {\it i.e.}, ${\rm
det}U=1$, because of ${\rm det}U=\exp [{\rm tr}(i{\mathcal G})]$.
Since it is known that such $U$ that satisfies simultaneously the
above two conditions is the group element of the su($n$) Lie
group, the high-dimensional Gell-Mann matrices, which closes the
corresponding su($n$) Lie algebraic commutation relations among
themselves, can also be constructed in terms of the
above-presented generators $a, a^{\dagger}, {\mathcal A},
{\mathcal M}, {\mathcal M}^{\dagger}, {\mathcal K}, {\mathcal F},
{\mathcal F}^{\dagger}, ...$. It is thus concluded that the P-B
harmonic oscillator with the maximum occupation number being $s$
has an $(s+1)$- dimensional number-state space and possesses the
su($s+1$) Lie algebraic structure.

Considering the case of $s\rightarrow \infty$ is of physically
typical interest. Apparently, it is seen that ${\mathcal A}$ tends
to a unit matrix ${\mathcal I}$, and the off-diagonal matrix
elements of all other generators except $a, a^{\dagger}$ approach
the zero matrices ${\mathcal O}$. This, therefore, means that the
P-B harmonic oscillator with infinite-dimensional state space just
corresponds to the Bosonic fields.

In the above we extend the non-semisimple algebra of harmonic
oscillator with infinite-dimensional state space to a semisimple
algebraic case, which can characterize the algebraic structures of
the P-B oscillator. In what follows we will consider a
generalization of P-B oscillator, {\it i.e.}, the so-called
supersymmetric Pegg-Barnett oscillator, which may possess some
physically interesting significance.

 For this aim, we will take into account a set of algebraic generators
$(N,N^{^{\prime }},Q,Q^{\dagger})$ which possesses a
supersymmetric Lie algebraic properties, {\it i.e.},

\begin{eqnarray}
Q^{2} &=&(Q^{\dagger })^{2}=0,\quad \left[ Q,Q^{\dagger}\right]
=N^{^{\prime }}\sigma _{z},\quad \left[ N,N^{^{\prime }}\right] =0,\quad %
\left[ N,Q\right] =-Q,  \nonumber \\
\left[ N,Q^{\dagger }\right] &=&Q^{\dagger },    \quad \left\{
Q,Q^{\dagger }\right\} =N^{^{\prime }},     \quad \left\{ Q,\sigma
_{z}\right\} =\left\{
Q^{\dagger },\sigma _{z}\right\} =0,  \nonumber \\
\left[ Q,\sigma _{z}\right] &=&-2Q,      \quad
 \left[ Q^{\dagger },\sigma _{z}
\right] =2Q^{\dagger },              \quad                 \left(
Q^{\dagger }-Q\right) ^{2}=-N^{^{\prime }},  \label{eq33}
\end{eqnarray}
where $\left\{ {}\right\} $ denotes the anticommuting bracket.
Such Lie algebra (\ref{eq33}) can be physically realized by the
two-level multiphoton Jaynes-Cummings model, the Hamiltonian
(under the rotating wave approximation) of which is of the
form\cite{Sukumar,Kien,Shen2}

\begin{equation}
H=\omega a^{\dagger }a+\frac{\omega _{0}}{2}\sigma
_{z}+g(a^{\dagger })^{k}\sigma _{-}+g^{\ast }a^{k}\sigma _{+},
\label{eq31}
\end{equation}
where $a^{\dagger }$ and $a$ are the creation and annihilation
operators for the electromagnetic field, and obey the commutation
relation $\left[ a,a^{\dagger }\right] =1$; $\sigma _{\pm }$ and
$\sigma _{z}$ denote the two-level atom operators which satisfy
the commutation relation $\left[ \sigma _{z},\sigma _{\pm }\right]
=\pm 2\sigma _{\pm }$ ; $g$ and $ g^{\ast }$ are the coupling
coefficients and $k$ is the photon number in each atom transition
process; $\omega _{0}$ and $\omega$ represent respectively the
transition frequency and the mode frequency. By the aid of
(\ref{eq33}) and the following expressions
(\ref{eq32})\cite{Lu1,Lu2,Shen1}
\begin{eqnarray}
N &=&a^{\dagger }a+\frac{k-1}{2}\sigma _{z}+\frac{1}{2}=\left(
{\begin{array}{cc}
{a^{\dagger }a+\frac{k}{2}} & {0} \\
{0} & {aa^{\dagger }-\frac{k}{2}}    \\
\end{array} }
\right) ,                             \quad N^{^{\prime }}=\left(
{\begin{array}{cc}
{\frac{a^{k}(a^{\dagger })^{k}}{k!} }& {0} \\
{0} & {\frac{(a^{\dagger })^{k}a^{k}}{k!} }
\end{array}  }
\right) ,  \nonumber \\
Q^{\dagger } &=&\frac{1}{\sqrt{k!}}(a^{\dagger })^{k}\sigma
_{-}=\left( {\begin{array}{cc}
{0} & {0} \\
{\frac{(a^{\dagger })^{k}}{\sqrt{k!}} }& {0}
\end{array}  }
\right) ,              \quad
 Q=\frac{1}{\sqrt{k!}}a^{k}\sigma
_{+}=\left( {\begin{array}{cc}
{0} & {\frac{a^{k}}{\sqrt{k!}} }\\
{0} & {0}          \\
\end{array}  }
\right) ,  \label{eq32}
\end{eqnarray}
the Hamiltonian (\ref{eq31}) of the two-level multiphoton
Jaynes-Cummings model can be rewritten as

\begin{equation}
H=\omega N+\frac{\omega -\delta }{2}\sigma _{z}+g Q^{\dagger
}+g^{\ast }Q-\frac{\omega}{2}         \label{eq34}
\end{equation}
with the detuning frequency $\delta =k\omega -\omega _{0}$.

The present illustrative example (the supersymmetric multiphoton
Jaynes-Cummings model) is helpful for understanding the physical
meanings of above supersymmetric algebra. But note that the
concept of supersymmetric P-B oscillator that will be put forward
in the following is not related to the above multiphoton
Jaynes-Cummings model (\ref{eq31}) at all. Use is made of
$\frac{1}{k!}a^{k}(a^{\dagger })^{k}\left| m\right\rangle
=\frac{(m+k)!}{m!k!}\left| m\right\rangle $ and
$\frac{1}{k!}(a^{\dagger })^{k}a^{k}\left| m+k\right\rangle
=\frac{(m+k)!}{m!k!}\left| m+k\right\rangle$, and then one can
arrive at
\begin{equation}
N^{^{\prime }} \left( {\begin{array}{*{20}c}
   {\left| m\right\rangle}  \\
   {\left| m+k\right\rangle}  \\
\end{array}} \right) =C_{m+k}^{m}\left( {\begin{array}{*{20}c}
   {\left| m\right\rangle}  \\
   {\left| m+k\right\rangle}  \\
\end{array}} \right)   \label{eq35}
\end{equation}
with $C_{m+k}^{m}=\frac{(m+k)!}{m!k!}.$ Thus we obtain the
supersymmetric quasialgebra $(N,Q,Q^{\dagger },\sigma _{z})$ in
the sub-Hilbert-space corresponding to the particular eigenvalue
$C_{m+k}^{m}$ of the Lewis-Riesenfeld invariant operator
$N^{^{\prime }}$\cite{Shen1} by replacing the generator
$N^{^{\prime }}$ with $C_{m+k}^{m}$ in the commutation relations
in (\ref{eq33}), namely,

\begin{equation}
\left[ Q,Q^{\dagger }\right] =C_{m+k}^{m}\sigma _{z}, \quad
\left\{ Q,Q^{\dagger }\right\} =C_{m+k}^{m},         \quad \left(
Q^{\dagger }-Q\right) ^{2}=-C_{m+k}^{m}.  \label{eq36}
\end{equation}

Based on the discussion of such quasialgebra in the
sub-Hilbert-space corresponding to the particular eigenvalue
$C_{m+k}^{m}$ of the generator $N^{^{\prime }}$, we can propose
the supersymmetric P-B oscillator. The algebraic generators of the
generalized P-B oscillator under consideration agree with the
commutation relation (\ref{eq36}), where $Q^{\dagger}$ and $Q$ can
be regarded as the creation and annihilation operators and the
eigenvalue $C_{m+k}^{m}$ of $N'$ may be considered the particle
occupation number of the P-B oscillator in a certain number state.
Evidently, if $k=0$, then the supersymmetric P-B oscillator is
reduced to the regular Fermionic case.

In addition, the Hamiltonian of the supersymmetric P-B oscillator
may be written in the form
\begin{equation}
H=\frac{1}{2}\left\{ Q,Q^{\dagger
}\right\}\Omega=\frac{1}{2}C_{m+k}^{m}\Omega
\end{equation}
by analogy with the Hamiltonian ($H=\frac{1}{2}\left\{a,a^{\dagger
}\right\}\omega$) of the Bosonic oscillator.

To summarize, in this note we consider the su($n$) Lie algebraic
structures in the P-B quantization formulation and generalize the
P-B oscillator to the supersymmetric case. It is shown that the
Fermionic and Bosonic fields are two special cases of P-B
oscillator, the corresponding dimensions of state spaces of which
are 2 and infinity, respectively. We think that some other related
topics in this supersymmetric case such as fractional statistics,
anyon\cite{Wilczek} and cyclic representation of quantum
algebra/group\cite{Fujikawa} may also deserve consideration.
\\ \\

\textbf{Acknowledgements}  This work was supported partially by
the National Natural Science Foundation of China under Project No.
$90101024$.

\end{document}